\documentclass[a4paper,12pt]{article}
\pdfoutput=1 
\usepackage{graphicx}
\usepackage{cite}
\usepackage{ytableau}
\usepackage[utf8]{inputenc}
\usepackage{tikz}
\usepackage{jheppub} 
\newcommand{\be}{\begin{equation}}
\newcommand{\eeq}{\end{equation}}
\newcommand{\bea}{\begin{eqnarray}}
\newcommand{\eea}{\end{eqnarray}}
\newcommand{\ba}{\begin{array}}
	\newcommand{\ea}{\end{array}}

\newcommand{\ee}{\end{equation} }

\newcommand{\tr}{\mathrm{tr}\,}

\newcommand{\one}{{\rm 1\kern -.9mm l}}

\title{VEV of $Q$-operator in $U(1)$ linear quiver 5d gauge theories }

\author{Gabriel Poghosyan}

\affiliation{Yerevan Physics Institute,\\
Alikhanian Br. 2, AM-0036 Yerevan, Armenia}
\emailAdd{gabrielpoghos@gmail.com}

\abstract{Linear quiver ${\cal N}=1$ 5d gauge theory in $\Omega$ background is 
	considered. It is shown that under certain restrictions on the VEV's of the 
	adjoint scalar field corresponding to the first node, only the array of Young 
	diagrams, such that the first diagram is a single column and the others are empty, 
	contribute to the partition function. Furthermore it is proved that this partition 
	function in a simple way is related to the expectation values of Baxter's $Q$ 
	operator (at specific discrete values of the spectral parameter) in the gauge 
	theory with the special node removed. Using known expression of the partition 
	function in the $U(1)$ quiver, Baxter's T-Q difference equations are established 
	and explicit expressions for the VEV of the $Q$ operator in terms of generalized 
	q-deformed Appel's functions is found. Finally the corresponding expressions for 
	the 4d limit are derived.         
}


\preprint{YerPhI/2018/01}
\begin{document}
	\maketitle
	\flushbottom
\section{Introduction}
	The 4d ${\cal N}=2$ gauge theories have natural uplift to the 5
	dimensions. Embedding ${\cal N}=2$ gauge theory in $\Omega$-background 
	was instrumental in all developments related to the instanton counting 
	with the help of equivariant localization technics. In fact the 
	geometric meaning of $\Omega$-background is more transparent in 5d 
	theory compactified on a circle. One simply considers a 5d geometry 
	fibered over a circle of circumference $L$ so that the complex 
	coordinates $(z_1,z_2)$ of the (four real dimensional) fiber get 
	rotated along the circle as: $z_1\rightarrow \exp (iL \epsilon_1)$, 
	$z_2\rightarrow \exp (iL \epsilon_2)$ accompanied with suitable 
	${\bf R}$-symmetry and gauge rotations \cite{Nekrasov:2002qd, Nekrasov:2003rj}. $\epsilon_{1,2}$ are the $Omega$-background 
	parameters. In 5d setting we'll use the notation 
	$T_{1,2}= \exp (-\beta \epsilon_{1,2})$, where $\beta=iL$ and 
	for technical reasons it will be assumed that $\beta$ has a tiny 
	real positive part.  The initial 4d theory 
	is recovered by sending $R\rightarrow 0$. Furthermore, sending both 
	$\Omega$-background parameters $ \epsilon_{1,2}$ to $0$, one gets the 
	standard Seiberg-Witten theory \cite{Seiberg:1994aj, Seiberg:1994rs}. 
	It is interesting that even the 
	case of $U(1)$ gauge group, in contrast to the case without
	$\Omega$-background, the theory is non-trivial. A characteristic feature 
	of this case, is that the instanton sums become tractable, and for 
	Nekrasov partition function one obtains closed formulae. In this paper 
	it is shown that not only the partition function, but also a more 
	refined quantity, namely the expectation value of the $Q$-observable 
	can be computed in closed form. It was shown in \cite{Poghossian:2010pn} 
	that the analog of Baxters $Q$ operator 
	in purely gauge theory context naturally emerges in Necrasov-Shatashvili 
	limit ($\epsilon_2=0$) \cite{Nekrasov:2009rc} as an entire function whose zeros are given in terms of an array of "critical" Young diagrams, namely  those, that determine the most important instanton configuration contributing to the partition function. This observable encodes perfectly not only information about partition function (which is simply related to the total sum of column lengths of Young diagrams) but also the entire chiral 
	ring \cite{Cachazo:2002ry} constructed from $\langle \tr \Phi^J\rangle $, 
	$J=0,1,2,\ldots $ ($\Phi$ is the scalar of vector multiplet) which can
	be expressed in terms of power sum  symmetric functions of the column
	lengths. This is why it is not surprising that the logarithmic 
	differential of (shifted) ratio $y(x)\sim Q(x)/Q(x+\epsilon_1)$ is the 
	direct analog of Seiberg-Witten differential: $x d \log(y(x)) \sim \omega_{SW}$. Subsequently $Q$ and $y$ observables have been extended for 
	theories with various matter and gauge contents \cite{Poghossian:2010pn, 
	Fucito:2011pn, Fucito:2012xc, Nekrasov:2013xda}. In particular \cite{Nekrasov:2013xda} interprets the equations satisfied by $y(x)$ as 
	deformed character relations and also considers the 5d setup, while in
	\cite{Poghossian:2016rzb} a relation between the $T-Q$ difference  
	equation and the  AGT dual \cite{Alday:2009aq,Wyllard:2009hg} (quasi-classical) 2d Toda conformal blocks with a fully degenerate insertion is found.  
	The next step, namely  extension to the case of generic
	$\Omega$-background has been achieved in \cite{Nekrasov:2015wsu} and \cite{Bourgine:2015szm}, where 
	Dyson-Shwinger type equations (called $qq$-character relations) for 
	$y$-observable are derived. For recent developments see also the series 
	of papers by Nekrasov\cite{Nekrasov:2015wsu,Nekrasov:2016qym, 
	Nekrasov:2016ydq,Nekrasov:2017rqy,Nekrasov:2017gzb}.

	In \cite{Poghosyan:2016mkh} the already mentioned link between 
	$Q$ observable and Toda conformal blocks with a degenerate field 
	insertion remains valid for the case of generic $\Omega$-background 
	and, in AGT dual 2d CFT side, fully quantum conformal blocks as well. 
	The case of the gauge group $SU(2)$ corresponding to the Liouville 
	theory was analyzed in much details and starting from 
	second order  the BPZ differential equation \cite{Belavin:1984vu}  a difference-differential equation, generalizing conventional Baxters 
	$T-Q$ relation \cite{Baxter:1982} was derived. In present paper simpler 
	$U(1)$ case in 5d setting is analyzed. The corresponding $T-Q$ difference 
	equations as well as their solutions in closed form are found.
	The solution is expressed in terms of generalized Appel's function. 
	
	The rest of material is organized as follows.\\
	In chapter \ref{5d_quiver} is a short review of 5d linear quiver gauge 
	theory: the Nekrasov partition function and important observables 
	$Q$, $y$ are introduced.\\
	In chapter \ref{Special_quiver_versus_Q} an extended quiver with 
	specific parameters at the extra nod is introduced and its relation to the 
	$Q$-observable is analyzed.\\
	Chapter \ref{eqs_solutions_5d} specializes to the case of $U(1)^r$ theory. 
	Difference equations $Q$-observable are derived. Explicit expressions 
	for the $Q$ observable in terms of generalized Appel and hypergeometric 
	functions are found.\\
	In chapter \ref{4d_reduction} through dimensional reduction, 
	corresponding difference equations and their solutions for the 4d 
	theory are found.\\
	In appendices \ref{A}, \ref{B}, \ref{C} some technical details, used in the main text, 
	are presented.

\section{5d linear quiver theory}\label{5d_quiver}
	\subsection{The partition function}
	The (instanton part of) partition function of the 5d, $A_{r+1}$ 
	linear quiver theory with gauge group $U(n)$ is given by (see 
	Fig.\ref{quiv_standard} where the setup and the notations are briefly described)
	\begin{gather}   
	\label{pf}
	\mathcal{Z}=\sum_{(\vec{Y}_1,...,\vec{Y}_r)}Z_{\textbf{Y}}
	q_1^{|\vec{Y}_1|}\dots q_r^{|\vec{Y}_r|} 
	\end{gather}
	The sum in (\ref{pf}) is  over all possible $r$-tuples of arrays of $n$ Young diagrams. 
	$|\vec{Y}_k|$ is the total number of boxes in the $k$-th array of $n$ Young 
	diagrams and $Z_{\textbf{Y}}$ is defined as:
	\begin{gather}  
	Z_{\textbf{Y}}=Z_{\vec{Y}_1,\dots,\vec{Y}_r}(\vec{a}_0,\vec{a}_1,\dots,\vec{a}_{r+1})
	=\nonumber\\\prod^{n}_{u,v=1}\frac{Z_{bf}(\scalebox{1.3}{\o},a_{0,u}|Y_{1,v},a_{1,v})
	Z_{bf}(Y_{1,u},a_{1,u}|Y_{2,v},a_{2,v})\dots 
	Z_{bf}(Y_{r,u},a_{r,u}|\scalebox{1.3}{\o},a_{r+1,v})}
	{Z_{bf}(Y_{1,u},a_{1,u}|Y_{1,v},a_{1,v})\dots 
	Z_{bf}(Y_{r,u},a_{r,u}|Y_{r,v},a_{r,v})} 
    \label{pf1}
	\end{gather}
	For a pair of Young diagrams $\lambda $, $\mu$  the bifundamental contribution 
	is given by \cite{Flume:2002az,Bruzzo:2002xf}
	\begin{gather} 
	\label{zbf}
	Z_{bf}(\lambda,a|\mu,b)=\prod_{s\in \lambda}\left(1-\frac{a}{b}
	T_{1}^{-L_{\mu}(s)}T_{2}^{1+A_{\lambda}(s)}\right)\prod_{s\in \mu}
	\left(1-\frac{a}{b}T_{1}^{1+L_{\lambda}(s)}T_{2}^{-A_{\mu}(s)}\right)
	\end{gather}
	$A_{\lambda}$ and $L_{\lambda}$, known as the arm and leg lengths respectively, 
	are defined as: if $s$ is a box with coordinates $(i,j)$ and 
	$\lambda_i$ ($\lambda_j'$) is the length of $i$-th ($j$-th) column (row), then:
	\begin{gather} 
	\label{alll}
	L_{\lambda}(s)=\lambda_j'-i\ , \quad A_{\lambda}(s)=\lambda_i-j
	\end{gather}
		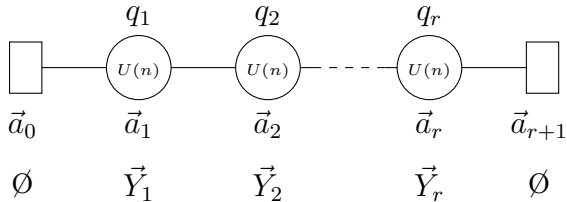
\begin{figure}[t]
		\begin{tikzpicture}[scale=0.85]
		\draw [] (0.5,0.1) rectangle (1,0.9);
		\node [below] at (0.7,0) {$\vec{a}_{0}$};
		\node [] at (0.7,-1.3) {$\scalebox{1.4}{\o}$};
		\draw [] (1,0.5)--(2,0.5);
		
		\draw [] (2.5,0.5) circle [radius=0.5];
		\node [] at (2.5,0.45) {\tiny{$U(n)$}};
		\node [below] at (2.5,0) {$\vec{a}_{1}$};
		\node [] at (2.5,-1.3) {$\vec{Y}_{1}$};
		\node [] at (2.5,1.3) {$q_{1}$};
		\draw [] (3,0.5)--(4,0.5);
		\draw [] (4.5,0.5) circle [radius=0.5];
		\node [] at (4.5,0.45) {\tiny{$U(n)$}};
		\node [below] at (4.5,0) {$\vec{a}_{2}$};
		\node [] at (4.5,-1.3) {$\vec{Y}_{2}$};
		\node [] at (4.5,1.3) {$q_{2}$};
		\draw [] (5,0.5)--(5.2,0.5);
		\draw [dashed] (5.2,0.5)--(6.3,0.5);
		\draw [] (6.3,0.5)--(6.5,0.5);
		\draw [] (7,0.5) circle [radius=0.5];
		\node [below] at (7,0) {$\vec{a}_{r}$};
		\node [] at (7,-1.3) {$\vec{Y}_{r}$};
		\node [] at (7,1.3) {$q_{r}$};
		\node [] at (7,0.45) {\tiny{$U(n)$}};
		\draw [] (7.5,0.5)--(8.5,0.5);
		\draw [] (8.5,0.1) rectangle (9,0.9);
		\node [below] at (8.7,0) {$\vec{a}_{r+1}$};
		\node [] at (8.7,-1.3) {$\scalebox{1.4}{\o}$};
		\end{tikzpicture}
		\caption{The linear quiver U(n) gauge theory: r
			circles stand for gauge multiplets; two squares represent n anti-fundamental 
			(on the left edge) and n fundamental (the right edge) matter multiplets while 
			the line segments connecting adjacent circles represent the bi-fundamentals. 
			$q_1,\dots, q_r$ are the exponentiated gauge couplings,
			the $n$-dimensional vectors $\vec{a}_{0},\dots ,
			\vec{a}_{r+1}$ encode respective (exponentiated)
			masses/VEV's and $\vec{Y}_{0},\dots ,
			\vec{Y}_{r+1}$ are $n$-tuples of young diagrams 
			specifying fixed (ideal) instanton configurations.} 
		\label{quiv_standard}
	\end{figure}
	\subsection{Important observables}
	The important observable of main interest in this paper, the $Q$-observable, 
	is defined as
	\begin{gather}\label{qd}
	{\bf Q}(x,\lambda)=\prod_{(i,j)\in \lambda}\frac{x-T_{1}^{i}T_{2}^{j-1}}
	{x-T_{1}^{i-1}T_{2}^{j-1}}
	\end{gather}
	Of course an analogous observable with the roles of $T_1$ and  $T_2$ 
	exchanged can be introduced as well. In 4d case $\beta\rightarrow 0 $ and in Nekrasov Shatashvili 
	limit $\epsilon_1\rightarrow 0 $ this observable
	satisfies Baxter's T-Q equation \cite{Poghossian:2010pn}: a difference 
	equation introduced by Baxter in context of lattice integrable models 
	\cite{Baxter:1982}. Generalization for the case of 
	generic $\Omega$ background (in both 4d and 5d cases) is due to \cite{Nekrasov:2015wsu}. 
	
	An important role is played also by the observable 
	\begin{gather}
	y(x,\lambda)=\frac{{\bf Q}(x,\lambda)}{{\bf Q}(x/T_2,\lambda)}\equiv 
	\prod_{(i,j)\in \lambda}\frac{(x-T_{1}^{i}T_{2}^{j-1})
		(x-T_{1}^{i-1}T_{2}^{j})}{(x-T_{1}^{i-1}T_{2}^{j-1})(x-T_{1}^{i}T_{2}^{j})}
	\end{gather}
	In 4d Nekrasov-Shatashvili limit the logarithmic derivative of 
	this observable generates all expectation values $\langle \phi^J \rangle$
	of the vector multiplet scalar. 
	Besides, its expectation value satisfies the (quantized analog of) 
	Seiberg-Witten curve equation \cite{Poghossian:2010pn}. In generic 
	$\Omega$-background the corresponding equations (the so called 
	qq-character equations) were 
	introduced and investigated in  \cite{Nekrasov:2015wsu} (see also \cite{Bourgine:2015szm}).
\section{The special quiver and its relation to the $Q$ observable}
\label{Special_quiver_versus_Q}
	The expectation value of the $Q$-operator associated to the first 
	node, by definition is 
	\begin{gather}
	\label{Q_def}
	Q(x)=Z^{-1}\sum_{(\vec{Y}_1,...,\vec{Y}_r)} 
	\prod_{u=1}^n{\bf Q}\left(\frac{x}{a_{1,u}},Y_{1,u}\right)
	Z_{\textbf{Y}}q_1^{|\vec{Y}_1|}\dots q_r^{|\vec{Y}_r|}
	\end{gather}
	It was noticed in \cite{Poghosyan:2016mkh} that such insertion of the 
	operator $Q$ is equivalent to 
	adding an extra node with specific expectation 
	values. Here this statement will be proved in more general 
	5d setting. Note that a detailed proof in \cite{Poghosyan:2016mkh} 
	was absent, so that also this gap automatically will be filled.
	   	
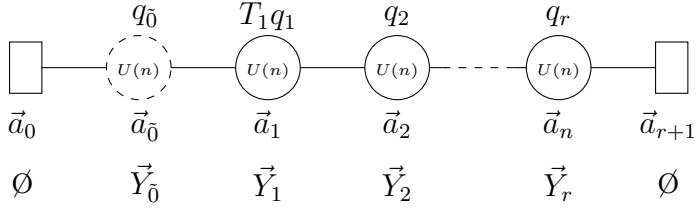
\begin{figure}[t]
	\begin{tikzpicture}[scale=0.85]
	\draw [] (0.5,0.1) rectangle (1,0.9);
	\node [below] at (0.7,0) {$\vec{a}_{0}$};
	\node [] at (0.7,-1.3) {$\scalebox{1.4}{\o}$};
	\draw [] (1,0.5)--(2,0.5);
	\draw [dashed] (2.5,0.5) circle [radius=0.5];
	\node [] at (2.6,1.3) {$q_{\tilde{0}}$};
	\node [] at (2.5,0.45) {\tiny{$U(n)$}};
	\node [below] at (2.6,0) {$\vec{a}_{\tilde{0}}$};
	\node [] at (2.6,-1.3) {$\vec{Y}_{\tilde{0}}$};
	\draw [] (3,0.5)--(4,0.5);
	\draw [] (4.5,0.5) circle [radius=0.5];
	\node [] at (4.5,0.45) {\tiny{$U(n)$}};
	\node [below] at (4.5,0) {$\vec{a}_{1}$};
	\node [] at (4.5,-1.3) {$\vec{Y}_{1}$};
	\node [] at (4.5,1.3) {$T_1q_{1}$};
	\draw [] (5,0.5)--(6,0.5);
	\draw [] (6.5,0.5) circle [radius=0.5];
	\node [] at (6.5,0.45) {\tiny{$U(n)$}};
	\node [below] at (6.5,0) {$\vec{a}_{2}$};
	\node [] at (6.5,-1.3) {$\vec{Y}_{2}$};
	\node [] at (6.5,1.3) {$q_{2}$};
	\draw [] (7,0.5)--(7.2,0.5);
	\draw [dashed] (7.2,0.5)--(8.3,0.5);
	\draw [] (8.3,0.5)--(8.5,0.5);
	\draw [] (9,0.5) circle [radius=0.5];
	\node [below] at (9,0) {$\vec{a}_{n}$};
	\node [] at (9,-1.3) {$\vec{Y}_{r}$};
	\node [] at (9,1.3) {$q_{r}$};
	\node [] at (9,0.45) {\tiny{$U(n)$}};
	\draw [] (9.5,0.5)--(10.5,0.5);
	\draw [] (10.5,0.1) rectangle (11,0.9);
	\node [below] at (10.7,0) {$\vec{a}_{r+1}$};
	\node [] at (10.7,-1.3) {$\scalebox{1.4}{\o}$};
	\end{tikzpicture}
	\caption{The quiver diagram with an extra node, labeled by $\tilde{0}$, added.
	Note that the gauge coupling at the node $1$ is chosen to be $T_1\,q_1$.}
	\label{quiv_extended}
\end{figure}

	Let's look at a quiver with $r+1$ nodes with expectation values at 
	the additional node (denoted as $\tilde{0}$) specified as 
	(see Fig.\ref{quiv_extended}):
	\begin{gather}
		a_{\tilde{0},u}=\frac{a_{0,u}}{T_{1}^{\delta_{1,u}}}.
	\end{gather}
	Due to the specific choice of $\vec{a}_{\tilde{0}}$, in order to give 
	a nonzero contribution, the array of $n$ diagrams associated with 
	the special node $\tilde{0}$ has to be 
	severely restricted. Namely, the diagram $Y_{\tilde{0},1}$ should 
	consist of a single column and the remaining $n-1$ diagrams 
	$Y_{\tilde{0},2}$, $\ldots$, $Y_{\tilde{0},n-1}$ must be empty. 
	The proof of this statement is given in the Appendix\,\ref{B}. 
	
	There is a close relation between the Nekrasov partition function 
	associated to above described specific length $r+1$ quiver and 
	the expectation value of a 
	particular $Q$ operator in a generic quiver with $r$ nodes.
	This relation is a consequence of the identity
	\begin{gather}
	Z_{\vec{Y}_{\tilde{0}},\vec{Y}_1,\dots,\vec{Y}_r}
	\left(\vec{a}_0,\vec{a}_{\tilde{0}},\vec{a}_1,\dots,
	\vec{a}_{r+1}\right)q^{l}_{\tilde{0}}
	\left(T_1 q_1\right)^{|\vec{Y}_1|}q_2^{|\vec{Y}_2|}\dots q_r^{|\vec{Y}_r|}
	= \nonumber \\ \prod_{u=1}^{n}\left({\bf Q}\left(\frac{a_{0,1}}{a_{1,u}}
	T^{l}_{2},Y_{1,u}\right) \frac{\left(\frac{a_{0,1}}{a_{1,u}}T_2;T_2\right)_l}
	{\left(\frac{a_{0,1}}{a_{0,u}}T_2;T_2\right)_l} \right)
	Z_{\vec{Y}_1,\dots,\vec{Y}_r}\left(\vec{a}_0,\vec{a}_1,
	\dots,\vec{a}_{r+1}\right)q^{l}_{\tilde{0}}q_1^{|\vec{Y}_1|}
	\dots q_r^{|\vec{Y}_r|}\,, 
	\label{qz}
	\end{gather}
    where $Y_{\tilde{0},u}$ for $u=1$ is a one column diagram with length $l$
    and the rest are empty diagrams. The q-analog of Pochhammer's symbol is defined as:
    \begin{gather}\label{ps}
    \left(a;q\right)_l=(1-a)(1-a q)\cdots\left(1-aq^{l-1}\right).
    \end{gather} 
	Inserting the definition (\ref{pf1}) of $Z_{\vec{Y}}$ and  
	canceling out the common factors of $q$ and $Z_{bf}$, we see that 
	(\ref{qz}) is equivalent to
	\begin{gather}
	\prod_{u,v=1}^{n}\left(	\frac{Z_{bf}(\scalebox{1.3}{\o},a_{0,u}|
	Y_{\tilde{0},v},a_{\tilde{0},v})Z_{bf}(Y_{\tilde{0},u},a_{\tilde{0},u}|
	Y_{1,v},a_{1,v})}{Z_{bf}(Y_{\tilde{0},u},a_{\tilde{0},u}|Y_{\tilde{0},v},
	a_{\tilde{0},v})}\right)T_{1}^{|\vec{Y}_1|}=\nonumber \\
	\prod_{u=1}^{n}\left({\bf Q}\left(\frac{a_{0,1}}{a_{1,u}}T^{l}_{2},
	Y_{1,u}\right)\frac{\left(\frac{a_{0,u}}{a_{1,u}}T_2,T_2\right)_l}
	{\left(\frac{a_{0,1}}{a_{0,u}}T_2;T_2\right)_l}\right)\prod_{u,v=1}^{n}
	Z_{bf}\left(\scalebox{1.3}{\o},a_{0,u}|Y_{1,v},a_{1,v}\right)
	\label{qz1}
	\end{gather}
	The last equality is proven in Appendix \ref{A}.
	
	Clearly, the eq. (\ref{qz}) shows that the VEV (\ref{Q_def}) at 
	specific values $x=x_l$
	\begin{gather}
	\label{xl}
	x_l=a_{0,1} T_2^l\,\,,\qquad\qquad l=0,1,2,\ldots
	\end{gather}
	is related to the 
	partition function of the special quiver with the fixed instanton number 
	$|\vec{Y}_{\tilde{0}}|=l$ at the node $\tilde{0}$
	\begin{eqnarray}
	\label{qz_Un}
	Q(x_l)&=&
	Z^{-1}\prod_{u=1}^{n} 
	\frac{\left(\frac{a_{0,1}}{a_{0,u}}T_2;T_2\right)_l} 
	{\left(\frac{a_{0,u}}{a_{1,u}}T_2;T_2\right)_l}\nonumber\\
	&&\times \sum_{(\vec{Y}_1,...,\vec{Y}_r)} 
	Z_{\vec{Y}_{\tilde{0}},\vec{Y}_1,\dots,\vec{Y}_r}
	\left(\vec{a}_0,\vec{a}_{\tilde{0}},\vec{a}_1,\dots,
	\vec{a}_{r+1}\right)\left(T_1 q_1\right)^{|\vec{Y}_1|}
	q_2^{|\vec{Y}_2|}\dots q_r^{|\vec{Y}_r|}
	\end{eqnarray}

\section{Difference equation for $Q$ and its solution}\label{eqs_solutions_5d} 
	From now on we'll restrict ourselves to the simplest case of the quiver of $U(1)$'s.
	5d Nekrasov partition function of such linear quiver can be found using refined topological 
	vertex method \cite{Bao:2011rc}, \cite{Mitev:2014jza}, 
	\cite{Iqbal:2007ii,Iqbal:2004ne,Taki:2007dh} or through a direct instanton 
	calculation (see e.g. \cite{Benvenuti:2016dcs} and references therein).
	The result can be represented as the infinite product
	\begin{gather}
	\label{ZU1}
	\mathcal{Z}=\prod_{l,s=0}^{\infty}\prod_{i=1}^{r}\prod_{j=i}^{r}
	\frac{\left(1-\frac{a_{i-1}p_i}{a_jp_j}\,T_{1}^{l}T_{2}^{s}\right)
	\left(1-\frac{a_{i}p_i}{a_{j+1}p_j}\,T_{1}^{l+1}T_{2}^{s+1}\right)}
	{\left(1-\frac{a_{i}p_i}{a_jp_j}\,T_{1}^{l}T_{2}^{s}\right)
	\left(1-\frac{a_{i-1}p_i}{a_{j+1}p_j}\,T_{1}^{l+1}T_{2}^{s+1}\right)}	
	\end{gather}  
	where 
	\begin{gather}
	p_i=a_1\prod_{l=1}^{i}q_l
	\end{gather} 
	
	Applying the formula (\ref{ZU1}) for the special quiver discussed in 
	Section \ref{Special_quiver_versus_Q}, and for brevity denoting 
	the partition function of the special quiver simply as $\mathcal{Z}(q_{\tilde{0}})$, 
	up to factors independent of $q_{\tilde{0}}$ we get
	\begin{gather}
	\mathcal{Z}(q_{\tilde{0}})\simeq\prod_{s=0}^{\infty}\prod_{i=0}^{r}	
	\frac{1-\frac{a_0p_i}{a_1a_{i+1}}\,q_{\tilde{0}}T_{1}^{1-\delta_{i,0}}
	T_{2}^{1+s}}
	{1-\frac{a_0p_i}{a_1a_{i}}\,q_{\tilde{0}}T_{2}^{s}}
	\label{q0_dep}
	\end{gather}
	Note now that in ratio $Z(q_{\tilde{0}})/Z(T_2^{-1}q_{\tilde{0}})$ 
	nearly all factors cancel out and one is lead to the relation
	\begin{gather}
	\mathcal{Z}(q_{\tilde{0}})\prod_{i=0}^{r}
	\left(1-\frac{a_0p_i}{a_1a_{i+1}}\,q_{\tilde{0}}
	T_{1}^{1-\delta_{i,0}}\right)=
	\mathcal{Z}(T_2^{-1}q_{\tilde{0}})\prod_{i=0}^{r}
	\left(1-\frac{a_0p_i}{a_1a_{i}}
	\,q_{\tilde{0}}T_{2}^{-1}\right)
	\end{gather}
	Expanding this equality in powers of $q_{\tilde{0}}$ and taking 
	into account (\ref{qz_Un}), we'll get a linear relation 
	(with rational in $x_l$ coefficients) among
	$r+2$ quantities $Q(x_l)$, $Q(x_l/T_2)$,\ldots, $Q(x_l/T_2^{r+1})$. 
	
	First let consider the simplest case $r=1$. An easy computation 
	allows us to establish the equality
	\begin{gather}
	\label{TQr1}
	Q\left(x\right)-\left(1+q_1\,
	\frac{a_1T_1x-a_0a_2}{a_2\left(x-a_1\right)}\right)
	Q\left(\frac{x}{T_2}\right)+q_1\,
	\frac{a_1\left(x-a_0T_2\right)\left(T_1x-a_2\right)}
	{a_2(x-a_1)(x-a_1T_2)}
	Q\left(\frac{x}{T_2^2}\right)=0
	\end{gather}
	which is valid for infinitely many values $x=x_l$, $l=0,1,2,\ldots$ 
	(see eq. (\ref{xl})). 
	
	An essential observation is in order here. Since $Q(x)$ and hence 
	the entire LHS of the eq. (\ref{TQr1}) restricted up to an arbitrary instanton 
	order is a rational function of $x$, the equality must be valid also 
	for generic values of $x$.
	It is not difficult to check that the $q$-hypergeometric function 
	(see Appendix C for definition) 
	\begin{eqnarray}
	\label{Qr1}
	Q(x)=\frac{\left(q_1;T_2\right)_\infty}{\left(
	\frac{q_1a_1T_1T_2}{a_2};T_2\right)_\infty}\,\,{{}_{2}\phi_1}
	\left({\frac{a_0}{x},\frac{a_1T_1T_2}{a_2}\atop 
	\frac{a_1}{x}};T_2,q_1\right)
\label{Q_rank1}
	\end{eqnarray}
	is a solution of (\ref{TQr1}). The ($x$-independent) normalization 
	coefficient in (\ref{Qr1}) is fixed from the asymptotic condition
	\begin{gather}
	\lim_{x\rightarrow\infty}Q(x)=1
	\label{Q_norm}
	\end{gather}
	In fact it is possible to argue that (\ref{Qr1}) is the only solution of (\ref{TQr1}) 
	with correct asymptotic and rationality properties discussed above. 
	Using the special $n=1$ case of the identity (\ref{iden_andrews}) the 
	eq. (\ref{Q_rank1}) can be rewritten also as (this equality is referred 
	as Heine's first transformation \cite{Gasper:2004:BHS})
\begin{eqnarray}
Q(x)=\frac{\left(\frac{a_0}{x};T_2\right)_\infty}
{\left(\frac{a_1}{x};T_2\right)_\infty}\,\,\,
{{}_{2}\phi_{1}}\left({\frac{a_1}{a_0},\ q_1,\ 
\atop \frac{q_1a_1T_1T_2}{a_{2}}};T_2,\frac{a_0}{x}\right).
\end{eqnarray}
	
	The general case with an arbitrary $r$ though more cumbersome, could 
	be analyzed in the same way. The resulting difference equation reads:
	\begin{gather}
	    \sum_{s=0}^{r+1}(-)^{s}C_s Q\left(T_{2}^{-s}x\right)=0
		\end{gather}
		where $C_s$
   \begin{gather}
   C_s=\frac{
   	xT_{1}^{s-1}(O_{+}^{(s-1)}+T_{1}O_{+}^{(s)})-a_{1}O^{(s-1)}
   	-a_0O^{(s)}}{x-a_0}\,\,\prod_{n=0}^{s-1}
   \frac{x-a_0T^{n}_{2}}{x-a_1 T_{2}^{n}}
   \end{gather}
   and $O^{(i)},O_{+}^{(i)}$ are the coefficients of the expansions:
   \begin{eqnarray}
   &&\prod_{i=1}^{r}\left(t+\frac{p_i}{a_i}\right)
   =\sum_{s=-\infty}^{\infty}O^{(s)}t^{r-s}\\
   &&\prod_{i=1}^{r}\left(t+\frac{p_i}{a_{i+1}}\right)=
   \sum_{s=-\infty}^{\infty}O_{+}^{(s)}t^{r-s}\,
   \end{eqnarray}
   or, explicitly 
    \begin{eqnarray}
    &&O^{(s)}=\sum_{1\le c_1<...<c_s\le r}\frac{p_{c_{1}} ... p_{c_{s}}}
    	{a_{c_{1}} ... a_{c_{s}}} \\
    &&O_{+}^{(s)}=\sum_{1\le c_1<...<c_s\le r}
    	\frac{p_{c_{1}} ... p_{c_{s}}}{a_{c_{1}+1} ... a_{c_{s}+1}}\,,
    \end{eqnarray}
    supplemented with conditions $O^{(-1)}=O_{+}^{(-1)}=0$ and 
     $O^{(0)}=O_{+}^{(0)}=1$.
     
	It is possible to find a closed expression for $Q$ by expanding the RHS of 
	eq. (\ref{q0_dep}) in powers of $q_{\tilde{0}}$ and, with the help of eq. 
	(\ref{qz_Un}), relating the coefficients to $Q(x_l)$. After few 
	manipulations, with a key role played by the identity 
	\begin{equation}
	\frac{(ax;q)_\infty}{(a;q)_\infty}=\sum_{l=0}^\infty
	\frac{(a;q)_l}{(q;q)_l}\,x^l\,,
	\label{q_bin}
	\end{equation}
	we finally get the expression 
    \begin{equation}
    \resizebox{.9\hsize}{!}{\(\displaystyle Q(x)=C\sum_{m_1,...,m_r\geq0}
    	\frac{\left(\frac{a_0}{x};T_2\right)_{m_1+m_2+...+m_r}
    		\left(\frac{a_1T_1T_2}{a_2};T_2\right)_{m_1}...
    		\left(\frac{a_rT_1T_2}{a_{r+1}};T_2\right)_{m_k}}
    	{\left(\frac{a_1}{x};T_2\right)_{m_1+m_2+...+m_r}(T_2;T_2)_{m_1}...
    		\left(T_2;T_2\right)_{m_r}}\,
    	\left(\frac{p_1}{a_1}\right)^{m_1}...\left(\frac{p_r}{a_r}\right)^{m_r}
    	,\)}
    \label{Q_msum}
    \end{equation}
    which is a generalization of the q-Appell's  $\Phi^{(1)}$ series. As 
    earlier, the normalization constant $C$ can be fixed from the 
    condition (\ref{Q_norm}). Indeed in large $x$ limit the RHS of eq.
    (\ref{Q_msum}) breaks down to $r$ independent sums which are easy 
    to evaluate using eq. (\ref{q_bin}). The end result is:
    \begin{equation}
    C=\prod_{i=1}^r\frac{\left(\frac{p_i}{a_i};T_2\right)_{\infty}}
    {\left(\frac{p_iT_1T_2}{a_{i+1}};T_2\right)_{\infty}}
    \end{equation}
    It is remarkable that the multiple sum (\ref{Q_msum}) can 
    be expressed in terms of basic hypergeometric series ${{}_{r+1}\phi_{r}}$ 
    (see Appendix \ref{C}) so that we finally get
\begin{eqnarray}
Q(x)=\frac{\left(\frac{a_0}{x};T_2\right)_\infty}
{\left(\frac{a_1}{x};T_2\right)_\infty}\,\,\,
{{}_{n+1}\phi_{n}}\left({\frac{a_1}{a_0},\ \frac{p_1}{a_1},\ 
\frac{p_2}{a_2},\ \dots\ ,\ \frac{p_r}{a_r}\atop 
\frac{p_1T_1T_2}{a_{2}},\frac{p_2T_1T_3}{a_{3}},%
\dots,\frac{p_rT_1T_2}{a_{r+1}}};T_2,\frac{a_0}{x}\right).
\label{Q_final}
\end{eqnarray}  
\section{Reduction to $4$ dimensions}\label{4d_reduction}
In this section we reduce our results to the case of four dimensions. We substitute:
\begin{gather}
 a_{i,u}\rightarrow e^{-\beta a_{i,u}}, \qquad T_1\rightarrow e^{-\beta \epsilon_1}, 
 \qquad T_2\rightarrow e^{-\beta \epsilon_2}.
\end{gather} 
where $a,\epsilon_1$ and $\epsilon_2$ are the parameters of our 4d quiver theory. 
The reduction corresponds to the small $\beta$ limit. Let me briefly list how the various 
quantities and relations get modified.\\
1. The link between expectation value of the $Q$ operator and the partition function:
\begin{eqnarray}
&&Z_{\vec{Y}_0,\vec{Y}_1,\dots,\vec{Y}_r}(\vec{a}_0,\vec{a}_{\tilde{0}},
\vec{a}_1,\dots,\vec{a}_{r+1})q^{l}_{0}q_1^{|\vec{Y}_1|}
q_2^{|\vec{Y}_2|}\dots q_r^{|\vec{Y}_r|}= \\ 
&&\prod_{u=1}^{n}\Big({\bf Q}(a_{0,1}-a_{1,u}+l\epsilon_2,Y_1)
\frac{(\frac{a_{0,u}-a_{1,u}+\epsilon_2}{\epsilon_2})_l}{(\frac{a_{0,1}-a_{0,u}
+\epsilon_2}{\epsilon_2})_l}\Big)Z_{\vec{Y}_1,\dots,\vec{Y}_r}
(\vec{a}_0,\vec{a}_1,\dots,\vec{a}_{r+1})q^{l}_{0}q_1^{|\vec{Y}_1|}\dots q_r^{|\vec{Y}_r|}
\nonumber
\end{eqnarray}
where $(a)_l$ is the standard Pochhammer's symbol, $Z_{\bf Y}$ is still given 
by eq. (\ref{pf1}) but now $Z_{bf}$ and $Q$ are given by
\begin{eqnarray}
\resizebox{.92\hsize}{!}{\(\displaystyle Z_{bf}(\lambda,a|\mu,b)=
\prod_{s\in\lambda}(a-b-L_{\mu}(s)\epsilon_1+(1+A_{\lambda}(s)\epsilon_2)
\prod_{s\in\mu}(a-b+(1+L_{\lambda}(s))\epsilon_1-A_{\mu}(s)\epsilon_2)\)}\qquad\\
{\bf Q}(x,\lambda)=\prod_{x\in\lambda}\frac{x-i\epsilon_1-(j-1)\epsilon_2}
{x-(i-1)\epsilon_1-(j-1)\epsilon_2}\hspace{7.2cm}
\end{eqnarray} 
2. The difference equation and its solution:
\begin{gather}
\sum_{s=0}^{r+1} (-)^s C_sQ(x-a_1-s\epsilon_2)=0
\end{gather}
where $C_s$ is defined as:
\begin{eqnarray}
\resizebox{.91\hsize}{!}{\(\displaystyle 
C_s=\frac{(a_1-x-(s-1)\epsilon_1)V^{(s-1)}+W^{(s-1)} +(a_0-x-s\epsilon_1)V^{(s)}+W^{(s)}}{a_0-x}\,
\prod_{n=0}^{s-1}\frac{a_0+n\epsilon_2-x}{a_1-m\epsilon_2-x}\)}\qquad
\end{eqnarray}
where
\begin{eqnarray}
V^{(i)}&=&\sum_{1\le c_1<...<c_i\le r}p_{c_{1}} ... p_{c_{i}} \\
W^{(i)}&=&\sum_{1\le c_1<...<c_i\le r}p_{c_{1}} ... p_{c_{i}}(a_{c_{1}+1}
-a_{c_{1}}+ ...+a_{c_{i}+1}-a_{c_{i}})\,.
\end{eqnarray}
Here $p_i$'s are redefined as:
\begin{equation}
p_i=\prod_{j=1}^{i}q_j
\end{equation}
and, by definition, we set $V^{(-1)}=W^{(-1)}=W^{(0)}=0$, $V^{(0)}=1$.\\
The solution reads: 
\begin{eqnarray}
Q(x)&=&C\sum_{m_1,...,m_r\geq0}\frac{(\frac{a_0-x}{\epsilon_2})_{m_1+m_2+...+m_r}
(\frac{a_1-a_2+\epsilon_1+\epsilon_2}{\epsilon_2})_{m_1}...
(\frac{a_r-a_{r+1}+\epsilon_1+\epsilon_2}{\epsilon_2})_{m_k}}
{(\frac{a_1-x}{\epsilon_2})_{m_1+m_2+...+m_r}m_1!...m_r!}\,
p_1^{m_1}...p_k^{m_r}\nonumber\\
&=&CF^{(r)}_1\big(\frac{a_0-x}{\epsilon_2},\frac{a_1-a_2+\epsilon_1+\epsilon_2}{\epsilon_2},...,
\frac{a_r-a_{r+1}+\epsilon_1+\epsilon_2}{\epsilon_2};\frac{a_1-x}{\epsilon_2};p_1,..,p_r\big)\,,\nonumber\\
\end{eqnarray}
where $F^{(r)}_1$ is a generalization of Appel's function (see Appendix \ref{C}). $C$ is 
$x$-independent and can be fixed from normalization:
\begin{eqnarray}
C=\prod_{i=1}^r\left(1-p_i\right)^{\frac{a_i-a_{i+1}+\epsilon_1+\epsilon_2}{\epsilon_2}}
\end{eqnarray}
In the special case $r=1$ we get
\begin{eqnarray}
Q(x)=\left(1-q\right)^{\frac{a_1-a_{2}+\epsilon_1+\epsilon_2}{\epsilon_2}}
{{}_{2}F_{1}}\left(\frac{a_0-x}{\epsilon_2},
\frac{a_1-a_2+\epsilon_1+\epsilon_2}{\epsilon_2};
\frac{a_1-x}{\epsilon_2};q\right)
\end{eqnarray}
It is not difficult to check that  after decoupling of both hypermultiplets 
by sending their masses to infinity, in Nekrasov-Shatashvili limit we 
recover the result presented in \cite{Poghossian:2010pn}.
\section{Acknowledgments}
This work was partially supported by the Armenian State Committee of Science 
in framework of the research projects 15T-1C233.
\appendix
\section{Proof of the equality (\ref{qz1})}\label{A}
Here we present the derivation of (\ref{qz1}).
We first derive two auxiliary identities.\\
Denote a Young diagram $\lambda$ with column lengths $\lambda_1\ge\lambda_2\ge \cdots $ as $\{\lambda_1,\lambda_2,\dots\}$. The corresponding row lengths we'll indicate as $\lambda'_1\ge\lambda'_2\ge \cdots $. In 
particular $\lambda'_1$ would be the number of columns. We want to show that
\begin{gather}
Z_{bf}(\{l\},a|\lambda,b)={\bf Q}\left(\frac{a}{b}
\,T_1T_2^l,\lambda\right)T_1^{-|\lambda|}
Z_{bf}(\scalebox{1.2}{\o},aT_1|\lambda,b)
\left(\frac{a}{b}T_1T_2;T_2\right)_l\label{aux1}
\end{gather} 
To prove (\ref{aux1}) we divide and multiply the LHS by 
$Z_{bf}(\scalebox{1.2}{\o},aT_1|\lambda,b)$, then insert the definitions of 
$Z_{bf}$ and the values of arm and leg lengths. We get
\begin{eqnarray}
\resizebox{.92\hsize}{!}{\(\displaystyle
Z_{bf}(\{l\},a|\lambda,b)=Z_{bf}(\scalebox{1.2}{\o},aT_1|\lambda,b) \prod_{j=1}^{l}\left(1-\frac{a}{b}T^{1-\lambda'_j}_{1}T^{1+l-j}_{2}\right)
\prod_{j=1}^{\lambda_1}\prod_{i=1}^{\lambda'_j}
\frac{1-\frac{a}{b}T^{1+\theta(l-j)-i}_{1}T^{-\lambda_i+j}_{2}}
{1-\frac{a}{b}T^{2-i}_{1}T^{-\lambda_i+j}}\)},\qquad 
\label{eq1}
\end{eqnarray}   
where $\theta(x)$ is the  Heaviside step function
\begin{eqnarray}
\theta(x)= \begin{cases} 1, & \mbox{if } x\geq 0 \\
 0, & \mbox{if } x<0 \end{cases}
\end{eqnarray} 
Now we divide the problem into two separate cases when  $\lambda_1\le l$ or 
$\lambda_1> l$.\\
1. $\lambda_1\leq l$. \\
In this case $\theta(l-j)=1$ and the double product in (\ref{eq1}) cancels 
out. The remaining single product by a simple manipulation can be rewritten 
as
\begin{eqnarray}
\resizebox{.92\hsize}{!}{\(\displaystyle
Z_{bf}(\scalebox{1.2}{\o},aT_1|\lambda,b) \prod_{j=\lambda_1+1}^{l}\left(1-\frac{a}{b}T^{1}_{1}T^{1+l-j}_{2}\right)
\prod_{j=1}^{\lambda_1}\prod_{i=1}^{\lambda'_j}
\frac{1-\frac{a}{b}T^{1-i}_{1}T^{1+l-j}_{2}}
{1-\frac{a}{b}T^{2-i}_{1}T^{1+l-j}_{2}}\prod_{j=1}^{\lambda_1}
\left(1-\frac{a}{b}T^{1}_{}T^{1+l-j}_{2}\right)\)}.\qquad
\end{eqnarray}
Notice that the middle double product is nothing but 
${\bf Q}(\frac{a}{b}\,T_1T_2^l,\lambda)T_1^{-|\lambda|}$ which concludes the first case.\\
2. $\lambda_1> l$ \\
We split $\lambda$ into two parts: $\lambda^{top}$ consisting of boxes with 
vertical coordinates $j>l$, and the part $\lambda^{down}$ of lower lying boxes 
with $j\leq l$. Now the part of the double product in (\ref{eq1}) corresponding
to the boxes of $\lambda^{top}$ survives. For the single product part we do 
the same manipulation as in previous case. As a result we get  
\begin{eqnarray}
\resizebox{.92\hsize}{!}{\(\displaystyle
Z_{bf}(\scalebox{1.2}{\o},aT_1|Y,b) \prod_{j=1}^{l}\prod_{i=1}^{\lambda'_j}\frac{1-\frac{a}{b}T^{1-i}_{1}
T^{1+l-j}_{2}}{1-\frac{a}{b}T^{2-i}_{1}T^{1+l-j}_{2}}
\prod_{j=1}^{l}(1-\frac{a}{b}T_1T^{1+l-j}_{2})\prod_{(i,j) \in \lambda^{top}}
\frac{1-\frac{a}{b}T^{1-i}_{1}T^{-\lambda_i+j}_{2}}
{1-\frac{a}{b}T^{2-i}_{1}T^{-\lambda_i+j}}\)}.\qquad 
\label{case_2} 
\end{eqnarray}
It is easy to see that the product over $\lambda^{top}$ can be rewritten as
\begin{eqnarray}
\prod_{(i,j) \in \lambda^{top}}
\frac{1-\frac{a}{b}T^{1-i}_{1}T^{-\lambda_i+j}_{2}}
{1-\frac{a}{b}T^{2-i}_{1}T^{-\lambda_i+j}}
=\prod_{(i,j) \in \lambda^{top}}\frac{1-\frac{a}{b}T^{1-i}_{1}T^{1+l-j}_{2}}
{1-\frac{a}{b}T^{2-i}_{1}T^{1+l-j}_{2}}.
\end{eqnarray}
Thus the first double product in (\ref{case_2}) (which is a product over 
the boxes of $\lambda^{down}$) naturally combines with that of over 
$\lambda^{top}$ to give a product over 
entire $\lambda$. As a result, instead of (\ref{case_2}) we may as well write
\begin{gather}
Z_{bf}(\scalebox{1.2}{\o},aT_1|\lambda,b)\left(\frac{a}{b}T_1T_2;T_2\right)_l 
\prod_{(i,j) \in \lambda} \frac{1-\frac{a}{b}T^{1-i}_{1}T^{1+l-j}_{2}}
{1-\frac{a}{b}T^{2-i}_{1}T^{1+l-j}_{2}}.
\end{gather}
As before, the product over $\lambda$ 
gives ${\bf Q}(\frac{a}{b}\,T_2^l,\lambda)T_1^{-|\lambda|}$, 
which concludes the proof of eq. (\ref{aux1}).

We'll need also the simple identity
\begin{eqnarray}
Z_{bf}(\scalebox{1.2}{\o},a|\lambda,b)=\prod_{(i,j)\in \lambda} 
\left(1-\frac{a}{b}\,T_1^{1-i}T_2^{1-j} \right)
\label{aux2}
\end{eqnarray}

Now the only thing that remains to be done is to make use of (\ref{aux1}) and 
(\ref{aux2}):
\begin{eqnarray}
\prod_{u,v=1}^{n}	\frac{Z_{bf}(\scalebox{1.1}{\o},a_{0,u}|Y_{\tilde{0},v},a_{\tilde{0},v})
Z_{bf}(Y_{\tilde{0},u},a_{\tilde{0},u}|Y_{1,v},a_{1,v})}
{Z_{bf}(Y_{\tilde{0},u},a_{\tilde{0},u}|Y_{\tilde{0},v},a_{\tilde{0},v})}
=\hspace{3.7cm}
\nonumber\\	
\prod_{u,v=2}^{n}	Z_{bf}(\scalebox{1.1}{\o},a_{0,u}|Y_{1,v},a_{1,v})
\prod_{u=2}^{n}\frac{Z_{bf}(\scalebox{1.1}{\o},a_{0,u}|Y_{1,1},a_{1,1})
Z_{bf}(Y_{0,1},\frac{a_{0,1}}{T_1}|Y_{1,u},a_{1,u})}
{Z_{bf}(Y_{0,1},\frac{a_{0,1}}{T_1}|\scalebox{1.1}{\o},a_{0,u})}\nonumber\\
\times\frac{Z_{bf}(\scalebox{1.1}{\o},a_{0,1}|Y_{0,1},
	\frac{a_{0,1}}{T_1})Z_{bf}(Y_{0,1},\frac{a_{0,1}}{T_1}|Y_{1,1},a_{1,1})}
{Z_{bf}(Y_{0,1},\frac{a_{0,1}}{T_1}|Y_{0,1},\frac{a_{0,1}}{T_1})} \nonumber \\
=
T_{1}^{-|\vec{Y}_1|}
\prod_{u=1}^{n}\Big({\bf Q}(\frac{a_{0,u}}{a_{1,u}}T^{l}_{2},Y_{1,u})
\frac{(\frac{a_{0,u}}{a_{1,u}}T_2;T_2)_l}{(\frac{a_{0,1}}{a_{0,u}}
T_2;T_2)_l}\Big)\prod_{u,v=1}^{n}Z_{bf}(\scalebox{1.1}{\o},a_{0,u}|
Y_{1,v},a_{1,v})
\end{eqnarray}
\section{Restriction on Young diagrams at the special node}\label{B}
To prove that the diagram $Y_{\tilde{0},1}$ at the special node $\tilde{0}$ should 
have at most one column in order to have a nonzero contribution to the partition  function,
let us assume in contrary that $Y_{\tilde{0},1}$ has a non-empty second column with length 
$l\ge 1$. This means 
that the box with coordinates $(i,j)=(2,l)$ belongs to this diagram. Any term 
of the instanton sum corresponding to such choice includes a factor   
\begin{gather}
Z_{bf}(\scalebox{1.2}{\o},a_{0,1}|Y_{\tilde{0},1},a_{0,1}T_1^{-1})
=\prod_{s\in Y_{\tilde{0},1}}\left(1-\frac{a_{0,1}}{a_{0,1}T_1^{-1}}\,
T_{1}^{1+L_{\o}(s)}
T_{2}^{-A_{ Y_{\tilde{0},1}}(s)}\right)\label{ocd}
\end{gather}
The arm and leg lengths of the box $(2,l)$ are easy to calculate: 
$L_{{\o}}(2,l)=-2$ and $A_{Y}(2,l)=0$ and the corresponding factor in eq. (\ref{ocd}) 
vanishes.\\
In a similar way we can easily argue that all remaining $n-1$ diagrams 
$Y_{\tilde{0},i}$, $i=2,\ldots,n$ must be empty. In fact, if any of this 
diagrams is non-empty (denote it as $\lambda$), then $Z_{\bf{Y}}$ will 
include a factor
\begin{gather}
Z_{bf}(\scalebox{1.3}{\o},a_{1,i}|\lambda,a_{1,i})=
\prod_{s\in \lambda}(1-T_{1}^{1+L_{\o}(s)}T_{2}^{-A_{\lambda}(s)})
\end{gather}
In this product the factor corresponding to the top box $(1,\lambda_1)$
of its 
first column becomes zero, since for this box $L_{{\o}}(1,\lambda_1)=-1$
and $A_{\lambda}(1,\lambda_1)=0$.\\
Thus we have proven that {\it at the special node} the first diagram has at most 
one column while the remaining diagrams are empty.
\section{Generalized Appel and hypergeometric functions}\label{C}
Appels functions and their q-analogues generalize ordinary hypergeometric and 
q-hypergeometric functions for the case with more than one arguments. Here are 
the definitions:
\begin{itemize}
\item{Appel's function  $F_1$ and its generalization for the arbitrary number 
of variables:}\end{itemize} 
\begin{eqnarray}
&&F_1(a,b_1,b_2;c;x,y)=\sum_{m,n=0}^{\infty}\frac{(a)_{m+n}(b_1)_m(b_2)_n}
{(c)_{m+n}m!n!}x^my^n\\
&&F^{(k)}_1(a,b_1,...,b_k;c;x_1,..,x_k)=\sum_{m_1,...,m_k\geq0}
\frac{(a)_{m_1+...+m_r}(b_1)_{m_1}...(b_k)_{m_k}}{(c)_{m_1+...+m_r}
m_1!...m_k!}(x_1)^{m_1}...(x_k)^{m_k}\nonumber\\
\end{eqnarray}	
\begin{itemize}
\item{The corresponding q-analogs:}\end{itemize}
\begin{eqnarray}
\Phi_1(a,b_1,b_2;c;q;x,y)=\sum_{m,n=0}^{\infty}\frac{(a;q)_{m+n}(b_1;q)_m(b_2;q)_n}
{(c;q)_{m+n}(q;q)_m(q;q)_n}x^my^n\hspace{4.33cm}
\end{eqnarray}
\begin{eqnarray}
\resizebox{.92\hsize}{!}{\(\displaystyle \Phi_1^{(n)}(a,b_1,b_2,...,b_n;c;q;x_1,..,x_n)
=\sum^{\infty}_{m_1,...,m_n=0}\frac{(a;q)_{m_1+...+m_r}
(b_1;q)_{m_1}...(b_n;q)_{m_n}}{(c;q)_{m_1+...+m_r}
(q;q)_{m_1}...(q;q)_{m_n}}x_1^{m_1}...x_n^{m_n}\)},\qquad
\label{q_Appel_gen}
\end{eqnarray}
\begin{itemize}
\item{The basic Hypergeometric function ${{}_{n+1}\phi_{n}}$:}\end{itemize}
\begin{eqnarray}
{{}_{n+1}\phi_n}\left({a_1,\ \dots\ ,\ a_{n+1}\atop 
b_1,
	\dots,b_n};T_2,x\right)=\sum_{m=0}^{\infty}\frac{(a_1;q)_m\cdots 
(a_{n+1};q)_m}{(q;q)_m(b_1;q)_m\cdots (b_n;q)_m}x^m
\end{eqnarray} 
There is a nice identity relating $\Phi_1^{(n)}$ with 
${{}_{n+1}\phi_n}$ (see \cite{Andrews1972}):
\begin{eqnarray}
\resizebox{.92\hsize}{!}{\(\displaystyle 
\Phi_1^{(n)}(a,b_1,b_2,...,b_n;c;q;x_1,..,x_n)=
\frac{\left(a,b_{1}x_{1},b_{2}x_{2},\dots,b_{n}x%
_{n};q\right)_{\infty}}{\left(c,x_{1},x_{2},\dots,x_{n};q\right)_{\infty}}\,{{}_
{n+1}\phi_{n}}\left({c/a,x_{1},x_{2},\dots,x_{n}\atop b_{1}x_{1},b_{2}x_{2},%
\dots,b_{n}x_{n}};q,a\right)\)},\quad
\label{iden_andrews}
\end{eqnarray}
where
\begin{eqnarray}
(a_1,a_2,\dots,a_k;q)_\infty=\prod_{l=1}^k(a_l;q)_\infty\,.
\end{eqnarray}
This identity allows us to rewrite the eq. (\ref{Q_msum}) in terms of the function ${{}_{n+1}\phi_{n}}$ (see eq. (\ref{Q_final})).
\providecommand{\href}[2]{#2}\begingroup\raggedright\endgroup
\end{document}